\documentclass[a4paper,10pt,reqno]{amsart}
\usepackage[margin=3cm]{geometry}
\usepackage{graphicx}
\usepackage{dsfont,amsmath,amssymb,color,units,pstricks}

\usepackage[first,bottomafter,timestamp]{draftcopy}

\newcommand \be  {\begin{equation}}
\newcommand \bea {\begin{eqnarray} \nonumber }
\newcommand \ee  {\end{equation}}
\newcommand \eea {\end{eqnarray}}

\title{Principal Regression Analysis and the index leverage effect}
\author{Pierre-Alain Reigneron \and Romain Allez \and Jean-Philippe~Bouchaud}
\address{Capital~Fund~Management, 6--8 boulevard Haussmann, 75009 Paris, France}
\urladdr{http://www.cfm.fr}
\date{\today}

\begin{document}

\begin{abstract}
We revisit the index leverage effect, that can be decomposed into a volatility effect and a correlation effect. We investigate the latter using a matrix regression analysis, 
that we call `Principal Regression Analysis' (PRA) and for which we provide some analytical (using Random Matrix Theory) and numerical benchmarks. 
We find that downward index trends increase the average correlation between stocks (as measured by the most negative eigenvalue of the conditional correlation matrix), and makes the market mode more uniform. 
Upward trends, on the other hand, also increase the average correlation between stocks but rotates the corresponding market mode {\it away} from uniformity. There are two time scales associated to
these effects, a short one on the order of a month (20 trading days), and a longer time scale on the order of a year. We also find indications of a leverage effect for sectorial correlations as well, 
which reveals itself in the second and third mode of the PRA.
\end{abstract}

\maketitle
\section{Introduction}

Among the best known stylized facts of financial markets lies the so-called ``leverage effect'' \cite{Glosten,leverage,egarch,Matacz,Perello1,Perello2}, a name coined by Black to describe the negative correlation between 
past price returns and future realized volatilities in stock markets \cite{Black}. \footnote{While this effect holds for most markets in developed economies, Tenenbaum et al. 
\cite{Tenenbaum} report that the situation appears to be  different for markets in developing countries.} It is indeed well documented that negative price returns induce increased future volatilities, 
an effect responsible for the observed skew on the implied volatility smile in stock option markets (see e.g. \cite{Bergomi1,Bergomi2,Ciliberti}). 

However, the association, made by Black, with a true leverage effect (i.e. that when the value of a stock goes down its debt to equity ratio increases, thereby making the company riskier and 
more volatile), is probably misleading. In particular, the amplitude of the leverage correlation for indices is noticeably stronger than for individual stocks, which even 
sounds paradoxical when the index return is by definition the average of individual stock returns! The volatility of an index in fact reflects both the volatility of underlying single stocks and
the average correlation between these stocks. The increased leverage effect for indices must therefore mean that both these quantities are sensitive to a downward move of the 
market. 

The aim of the present paper is to investigate more specifically this ``correlation leverage effect'', and make precise the common lore according to which correlations ``jump to one'' 
in crisis periods (see \cite{Erb,Solnik,Solnik2,Ramchand} for early studies of the time evolution of the correlations in financial markets).
Similar studies have appeared recently. In \cite{Ingve}, a careful study of the average correlation between stock returns during contemporaneous upward/downward trends of the market index has
confirmed that correlations are indeed stronger when the market goes down \cite{Lisanew}. Our analyses confirm and make more precise these results, first by extending them to different markets, 
and second by devising and exploiting a new tool to investigate conditional correlations, that we call ``principal regression analysis'' (PRA). 
The idea here is to regress the instantaneous correlation matrix on the 
value of the index return (or any other conditioning variable). While the intercept of the regression gives the average correlation matrix, the regression slopes define a second symmetric (but 
not definite positive) matrix that can be diagonalized, leading to modes (eigenvectors) of sensitivity to the conditioning variable(s). The interpretation of these eigenvectors 
is particularly transparent when they coincide with those of the correlation matrix itself. The corresponding eigenvalues quantify how the whole correlation structure of stock returns is affected 
by the conditioning variable. 
The nice point about the PRA is that Random Matrix Theory (RMT) provides, as for standard PCA, a useful guide to decide whether or not  
these sensitivity modes are statistically meaningful (for a review on RMT, see \cite{RMT-review}). When the conditioning variable is the past values of the index return, the conclusion of PRA 
is that the dominant mode is the market mode, 
associated to a negative eigenvalue, indeed corresponding to a correlation leverage effect. We characterize the temporal decay of this effect. Upon separating positive and negative index returns, 
we furthermore find that the correlation leverage effect is strongly asymmetric: whereas negative returns increase both the volatility of the underlying stocks and the average correlation 
between stocks, positive returns have weaker influence on these quantities (see Fig. 6 below). We furthermore find indications of a leverage effect for sectorial correlations as well, 
which reveals itself in the second and third modes of the PRA.

\section{Data, notations and definitions} 

We have considered 6 pools of stocks corresponding to 6 major stock indices: SP500, BE500, Nikkei, FTSE, CAC 40 and DAX. We analyze the daily returns in a time period spanning from
$01/01/2000$ to $04/26/2010$. Stocks are labelled by $\alpha=1,\dots,N$ (where $N$ depends on the market), and days by $t=1, \dots, T$ (where $T=2594$). 
Time average will be denoted by $\langle.\rangle$.
The return of stock $\alpha$ between the close of day $t-1$ and the close of day $t$ is denoted as $\eta_\alpha(t)$. We in fact understand $\eta_\alpha(t)$ as the {\it demeaned} return 
over the whole time period $T$. We define an inverse volatility weighted index return at time $t$ as:
\be
I(t) = \frac1N \sum_{\alpha=1}^N \widehat \eta_\alpha(t), \qquad \widehat \eta_\alpha(t) \equiv \frac{\eta_\alpha(t)}{\sigma_\alpha},
\ee
where $\sigma_\alpha$ is the average volatility of the stock $\alpha$ over the whole time period:
\be
\sigma_\alpha^2 := \frac1T \sum_{t=1}^T \eta_\alpha(t)^2.
\ee
We will further define the average instantaneous stock volatility $\sigma(t)$ at time $t$ as:
\be
\sigma(t)^2 := \frac1N \sum_{\alpha=1}^N \widehat \eta_\alpha(t)^2
\ee
while the average instantaneous correlation between all pairs of stocks $\rho(t)$ is defined as:
\be
\rho(t) := \frac{1}{N(N-1)} \sum_{\alpha \neq \beta=1}^N \frac{\widehat \eta_\alpha(t)\widehat \eta_\beta(t)}{\sigma(t)^2}.
\ee
The average over time of the above two quantities will be denoted as $\sigma_0^2$ and $\rho_0$.

The squared index return $I(t)^2$ is a rough proxy for the instantaneous index volatility. Using the above definitions and the fact that $N$ is large, it is easy to check that:
\be
I(t)^2 \approx \rho(t) \sigma(t)^2 + O(\frac1N),
\ee
showing that both the average stock volatility and the average correlation contribute to the index volatility. It is therefore natural to decompose the full index leverage effect in two 
contributions: one coming from the dependence of the average stock volatility on the past returns of the index, and a second one describing the average correlation. We thus define a 
full leverage correlation function ${\mathcal L}_I(\tau)$:
\be
{\mathcal L}_I(\tau) = \frac{\left\langle I(t-\tau) I(t)^2 \right\rangle}{\left\langle I(t)^2\right\rangle},
\ee
and two partial leverage correlation functions:
\be
{\mathcal L}_\sigma(\tau) = \frac{\left\langle I(t-\tau) \sigma(t)^2 \right\rangle}{\left\langle I(t)^2\right\rangle}, \qquad 
{\mathcal L}_\rho(\tau) = \frac{\left\langle I(t-\tau) \rho(t) \right\rangle}{\left\langle I(t)^2\right\rangle}.
\ee
All the above leverage correlation functions are normalized to be the regression slope of the corresponding observables on the past value of the index return, for example:
\be\label{regress}
\rho(t) = \rho_0 + {\mathcal L}_\rho(\tau) I(t-\tau) + \varepsilon(t,\tau),
\ee
where $\varepsilon(t,\tau)$ is some noise. (Remember that by construction, $I(t)$ has zero mean.)

In the limit of weak correlations, the two effects are additive and one should find:
\be\label{sum}
{\mathcal L}_I(\tau) \approx \rho_0 {\mathcal L}_\sigma(\tau) + \sigma_0^2 {\mathcal L}_\rho(\tau),
\ee
eliciting the contribution of the average stock volatility and of the average correlation to the full leverage correlation. The second term is 
responsible for the enhanced leverage effect for indices compared to single stocks.
\begin{figure}[h!btp] 
		\center
		\includegraphics[scale=0.45]{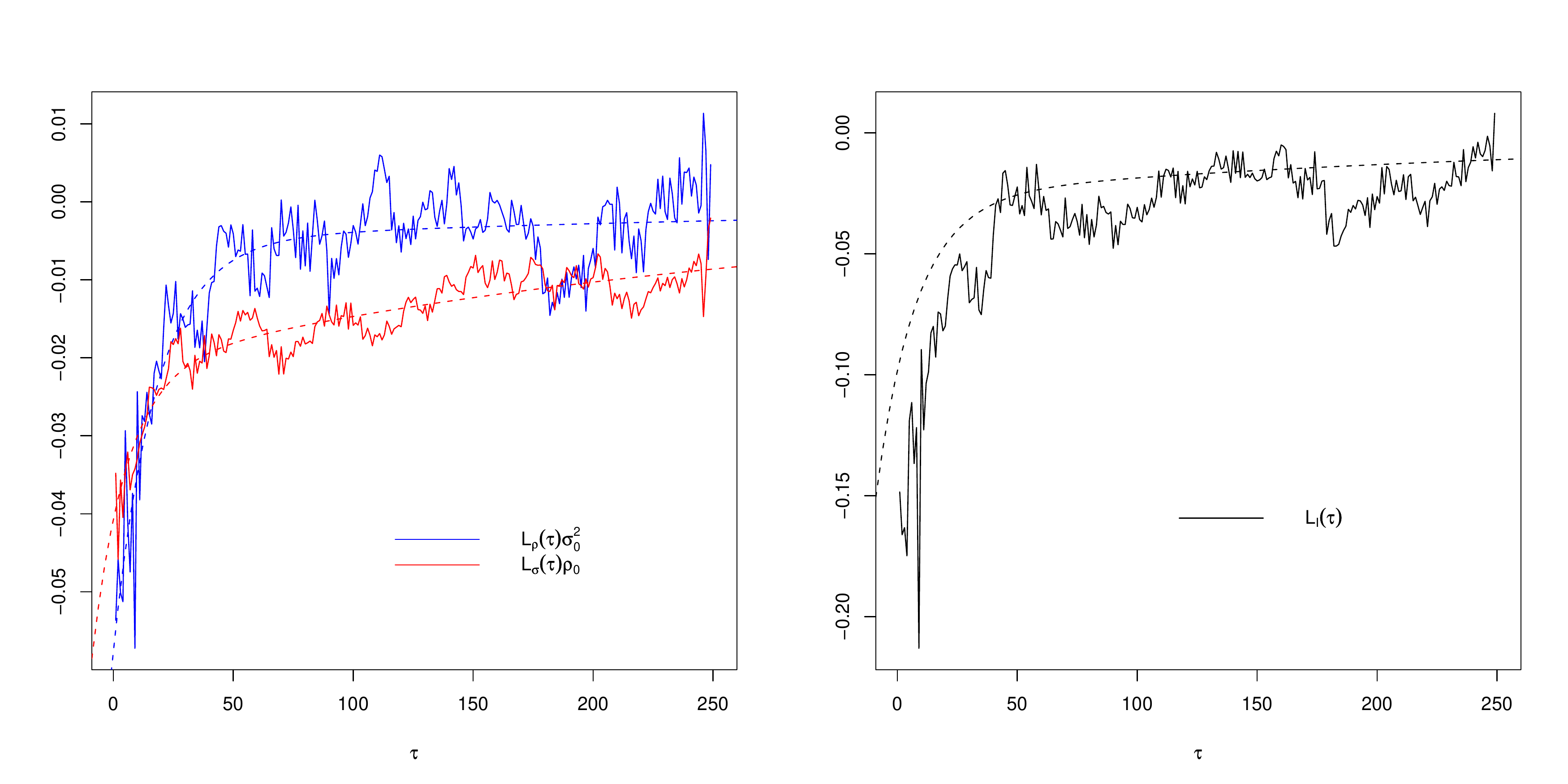}
        \caption{Left: normalized leverage correlation functions $\rho_0 {\mathcal L}_\sigma(\tau)$, 
        $\sigma_0^2 {\mathcal L}_\rho(\tau)$, and an exponential fits with two scales (dotted lines). 
        Right: Full 
        leverage function ${\mathcal L}_I(\tau)$ and comparison with an additive model (dotted line).} \label{Figure1}
\end{figure}

\section{Index leverage effect: A simple empirical analysis}

As a first stab at understanding the index leverage effect, we plot in Fig.~\ref{Figure1} the normalized partial leverage correlation functions, $\rho_0 {\mathcal L}_\sigma(\tau)$, $\sigma_0^2 {\mathcal L}_\rho(\tau)$, 
together with the full leverage ${\mathcal L}_I(\tau)$. In these plots, the data is averaged over the four indices, SP500, BE500, Nikkei and FTSE. From this figure, we draw the 
following conclusions: 
\begin{itemize} 
\item (a) the two contributions to the index leverage are of the same order of magnitude. In particular, the correlation leverage is significant and confirms the conclusions of Refs.~\cite{Ingve,Lisanew}.
\item (b) the correlation effect is stronger at short times but decays faster than the volatility effect; a two time scale exponential fit of these two contributions in the range 
$\tau \in [1,250]$ (in days) indeed leads to
\begin{align}
\sigma_0^2 {\mathcal L}_\rho(\tau) &\approx -0.053 \exp(-\tau/18) - 0.005 \exp(-\tau/350) ; \\
\qquad \rho_0 {\mathcal L}_\sigma(\tau)  &\approx -0.02 \exp(-\tau/14) - 0.02 \exp(-\tau/280),
\end{align}
\item (c) a test of Eq.~(\ref{sum}) with the sum of the above two fitted exponentials reproduces satisfactorily the full leverage effect, although the latter is underestimated at short times, when the
correlations cease to be small enough for Eq. (\ref{sum}) to be accurate.
\end{itemize} 

\begin{figure}[h!btp] 
		\center
		\includegraphics[scale=0.45]{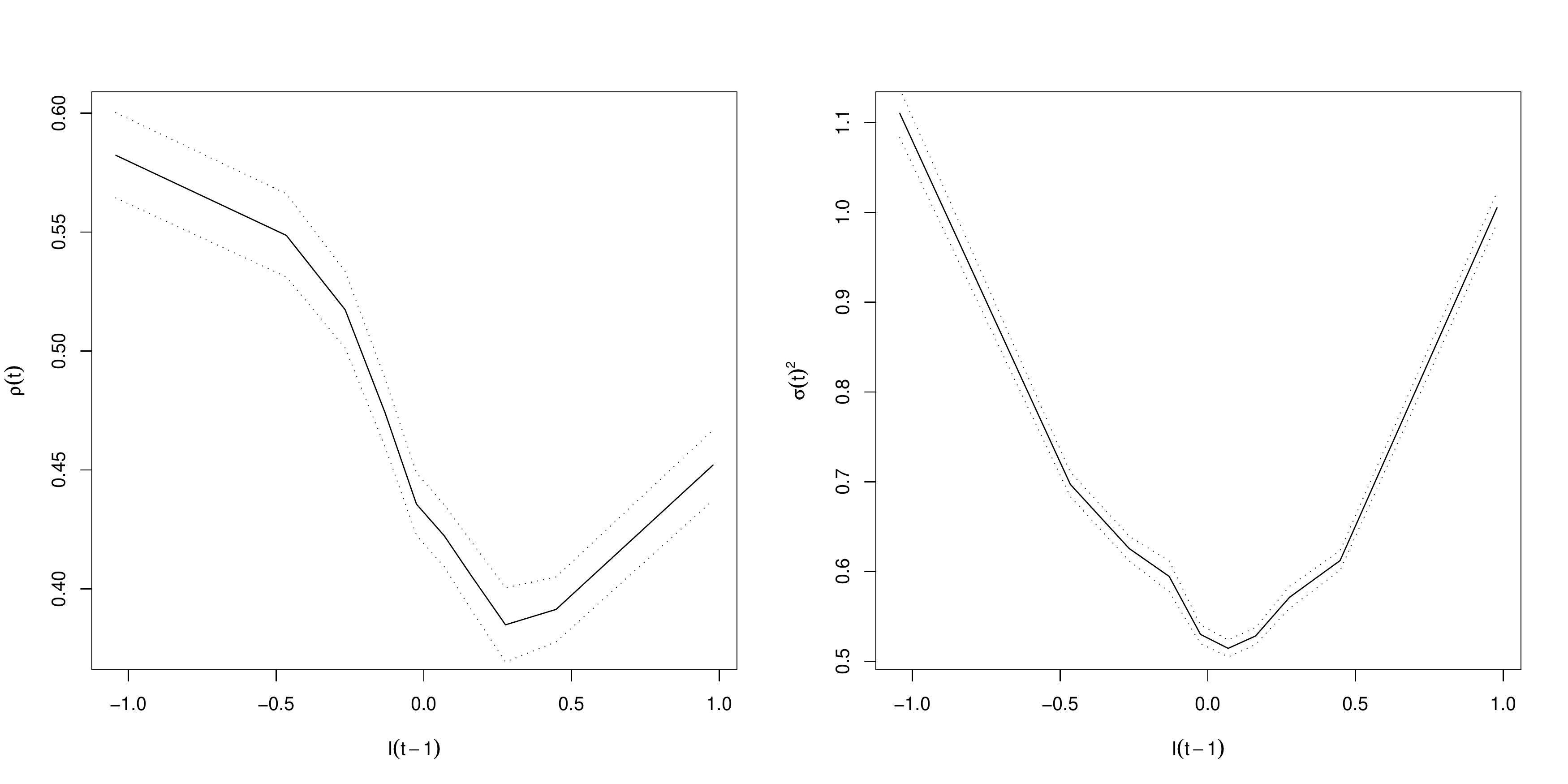}
        \caption{Dependence of the average correlation $\rho(t)$ and the average single stock volatility $\sigma^2(t)$ on the index return the previous day, $I(t-1)$. The result 
        is obtained as an average over all 6 indices: SP500, BE500, Nikkei, FTSE, CAC40 and DAX, but the qualitative effects are robust and appear on each markets individually. 
        These plots suggest that a quadratic $I^2(t-1)$ term should be included to the linear regressions. The printed error bars are the average of the error bars obtained for each of the $6$ indices.
        } \label{figure2}
\end{figure}

In fact, one can test directly whether linear regressions such as Eq.~(\ref{regress}) above make sense or not, by averaging 
all values of $\rho(t)$ corresponding to a given value of $I(t-1)$ within some range. The resulting graphs are shown in Fig.~\ref{figure2}, both for $\rho$ and for $\sigma^2$. One sees that whereas a linear
regression for $\rho$ makes sense for $I(t-1) < 0$, there is in fact perhaps a small positive slope for $I(t-1) > 0$. For $\sigma^2$, the graph looks even more symmetric, reflecting the presence 
of volatility correlations on top of (assymetric) leverage correlations.

\section{A more precise tool: The ``Principal Regression Analysis''} 

The above analysis, although interesting, is oversimplified, because the structure of inter-stock correlations is described by a full correlation matrix ${\bf C}$ and not by a single 
number $\rho$, that only captures the average correlations. In order to characterize the way the correlation matrix depends on the past value of the index (or on any other conditioning variable), we
propose the following: consider a given pair of stocks, $\alpha, \beta$, and regress the product of normalized returns $\widehat \eta_\alpha(t)\widehat \eta_\beta(t)$ on the past value of the index
return, i.e. write:
\be \label{defD}
\widehat \eta_\alpha(t)\widehat \eta_\beta(t) := C_{\alpha,\beta} + D_{\alpha, \beta}(\tau) I(t-\tau) + \varepsilon_{\alpha,\beta}(t,\tau).
\ee
Since $I(t)$ has zero mean, the intercept of the regression is exactly the empirical Pearson estimate of the correlation matrix. The regression {\it slopes} $D_{\alpha, \beta}(\tau)$ define 
another $N \times N$ symmetric matrix ${\bf D}(\tau)$, which encodes the full information about the dependence of the correlations on past returns. More precisely, the regression leads to the following empirical determination of ${\bf D}(\tau)$:
\be\label{empirical}
\langle I^2 \rangle D_{\alpha, \beta}(\tau) = \frac{1}{(T-\tau)} \sum_{t=\tau+1}^T \widehat \eta_\alpha(t) \widehat \eta_\beta(t) I(t-\tau).
\ee
The aim of this section is first to discuss the information 
contained in ${\bf D}(\tau)$, in particular its eigenvalues and eigenvectors, and second to use results from Random Matrix Theory to assess how meaningful this information is when the length 
of the sample, $T$, is not very large compared to the number of stocks $N$. Finally, we describe our empirical results on ${\bf D}(\tau)$, in particular its most negative eigenvalue and eigenvectors.

\subsection{Interpretation}

Define ${\bf C}(I)$ to be the correlation matrix conditioned to a certain past value of $I$, by:
\be
{\bf C}(I) = {\bf C} + I {\bf D}.
\ee

The interpretation of the matrix ${\bf D}$ is particularly simple when it commutes with the correlation matrix ${\bf C}$, i.e. when the eigenvectors of ${\bf D}$ are the same as those of
${\bf C}$. In this case, the eigenvectors of ${\bf C}(I)$ are exactly the same as those of ${\bf C}$, whereas the eigenvalues $\lambda_k(I)$ are shifted as:\footnote{Note that the dependence on the lag $\tau$ is implied in the following formulas.}
\be\label{pert1}
\lambda_k(I) = \lambda_k(0) + I \langle v_k | {\bf D} | v_k \rangle,
\ee
where $\lambda_k(0)$ are the eigenvalues of ${\bf C}$ and $| v_k \rangle$ are the associated eigenvectors (in quantum mechanics notations). When ${\bf D}$ does not commute with ${\bf C}$, the structure of the eigenvectors themselves is impacted by the conditioning variable. 
If ${\bf D} I$ is small enough, standard first order perturbation theory gives back Eq. (\ref{pert1}) for the eigenvalues and:
\be\label{pert2}
| v_k(I) \rangle = | v_k \rangle + I \sum_{\ell \neq k} \frac{\langle  v_\ell | {\bf D} | v_k \rangle}{\lambda_k-\lambda_\ell} | v_\ell \rangle,
\ee
for the eigenvectors of the matrix ${\bf{C}}(I)$. 

As we will find below, the eigenvector corresponding to the most negative eigenvalue of ${\bf D}$ turns out to be very close to the  first eigenvector of ${\bf C}$ (i.e. the so-called market mode, $| v_1 \rangle$), 
whereas all other eigenvalues are significantly smaller. In this case, the top eigenvalue of ${\bf C}$ is to a good approximation given by:
\be\label{levcorr}
\lambda_1(I) \approx \lambda_1 + I \mu_1,
\ee
where $\mu_1$ is the most negative eigenvalue of ${\bf D}$. Since $\lambda_1$ can be used to define the average correlation between stocks through $\lambda_1 := N \rho$, 
the meaning of $\mu_1$ is similar to, but more precise than, the correlation leverage
function ${\mathcal L}_\rho$ defined above.

More generally, when ${\bf D}$ and ${\bf C}$ do not commute, one expects the ``correlation leverage'' to rotate the top eigenvector away from the market mode $| v_1 \rangle$. 
The common lore is indeed that when markets go down, all stocks ``move together'', meaning that the top eigenvector should rotate towards the uniform vector  
$| e \rangle = (1/\sqrt{N}, 1/\sqrt{N}, \dots , 1/\sqrt{N})$. The cosine of the angle between $| v_1 \rangle$ and $| e \rangle$ is given by the scalar 
product $\langle e | v_1 \rangle$, that one can compute using perturbation theory. Eq.~(\ref{pert2}) above. Assuming further that the top eigenvalue of ${\bf C}$ is much larger than all the others ($\lambda_1 \gg \lambda_{\ell \neq 1}$), one finds:
\be\label{pert3}
\langle e | v_1(I) \rangle \approx \langle e | v_1 \rangle + \frac{I}{\lambda_1}  \left[\langle e | {\bf D} | v_1 \rangle -  \langle v_1 | {\bf D} | v_1 \rangle \langle e | v_1 \rangle \right].
\ee
A measure of how strongly the top eigenvector moves towards $| e \rangle$ is therefore provided by the quantity $\Delta$, defined as:
\be\label{delta}
\Delta = \frac{1}{\lambda_1} \left[\langle e | {\bf D} | v_1 \rangle -  \langle v_1 | {\bf D} | v_1 \rangle \langle e | v_1 \rangle \right].
\ee
A negative $\Delta$ means that the instantaneous market mode is closer to the uniform mode  $| e \rangle$ when the index goes down, since $\langle e | v_1(I) \rangle - \langle e | v_1 \rangle
= I \Delta > 0$.

\subsection{Results from Random Matrix Theory}

When $N$ is large, the simultaneous determination -- using Eq. (\ref{empirical}) above -- of the $N(N+1)/2$ different elements of ${\bf D}$ from the $NT$ data points is problematic, 
exactly in the same way the correlation matrix ${\bf C}$ 
is hard to measure. We thus need to provide a benchmark to compare the empirical results obtained with the noise level of the benchmark case. This will enable to separate significant effect from
noise level arising from the dimensionality problem.
Let $\xi$ be a random variable which will play the role of the conditioning variable (the past values of index returns in our context) and let $x_\alpha, \alpha = 1, \dots, N$ be a gaussian vector of covariance 
matrix $\bf C$ which should be seen as instantaneous stock
returns. The $x_\alpha$ will be supposed to have $0$ mean and unit variance, so that $\bf{C}$ is the correlation matrix of the gaussian vector $(x_1, \dots, x_N)$.

We begin by the case ${\bf C} = \bf I$.
Suppose, in addition, that there is {\it no correlations whatsoever} between the conditioning variable $\xi$ and the correlation $x_\alpha x_\beta$, and 
that one forms a matrix $\widetilde {\bf D}$ from:
\be\label{theoretical}
\langle \xi^2 \rangle \widetilde{D}_{\alpha, \beta} = \frac{1}{T} \sum_{t=1}^T x_\alpha(t) x_\beta(t) \xi(t).
\ee
In the limit 
$T \to \infty$ for finite $N$ one should find that all the elements of the matrix $\widetilde {\bf D}$ are zero, and therefore all its eigenvalues are zero as well. For finite $T$, however, the matrix $\widetilde {\bf D}$ will have a set of non trivial eigenvalues. Random Matrix Theory offers a way to compute the statistics of these eigenvalues when $N$ and $T$ are both large, with a fixed ratio $q=N/T$. The result depends both on the eigenvalue spectrum of the matrix ${\bf C}$ and, perhaps surprisingly, on the probability distribution of the conditionning variable, $P(\xi)$. The simplest, albeit unrealistic case for applications in finance, is when ${\bf C}$ is the identity matrix, i.e. there is no correlations between the $\widehat \eta$. In this case, using the theory of Free Random Matrices \cite{Verdu}, one finds that the empirical eigenvalue spectrum of $\widetilde {\bf D}$, 
$\rho_1(\mu)$, is the solution of the following set of equations, in the limit where $\epsilon$ goes to zero: \cite{RMT-review,Biroli}
\begin{eqnarray}\label{doseq1}
\mu&=&\frac{G_R}{G_R^2+\pi^2 \rho_1^2}+\int {\rm d}\xi P(\xi) \frac{\xi (1-q \xi G_R)}{(1-q \xi G_R)^2+(q\pi \xi\rho_1)^2}\\
\epsilon&=&\rho_1\left(-\frac{1}{G_R^2+\pi^2\rho_1^2}+\int {\rm d}\xi P(\xi) \frac{q\xi^2}{(1-q \xi G_R)^2+(q\pi \xi \rho_1)^2} \right),\label{doseq2}
\end{eqnarray}
where $G_R$ is the real part of the resolvent. One can check that in the limit $q \to 0$, and using the fact that $\xi$ has zero mean, the above equations boil down to:
\be
\frac{1}{G_R - i \pi \rho_1} = \mu - i \epsilon \to \rho_1(\mu) = \delta(\mu),
\ee
i.e. all eigenvalues are zero, as they indeed should when $T \gg N$. 

The case of an arbitrary correlation matrix ${\bf C}$ can also be solved completely using the above result on $\rho_1$ and the so-called $S$-transform of the eigenvalue spectrum \cite{Verdu}, 
noting that the eigenvalues of $\widetilde {\bf D}$ are the same as those of the product ${\bf C} \times {\bf D}_1$, where ${\bf D}_1$ is a random matrix with eigenvalue spectrum $\rho_1(\mu)$. 
The resulting equation can in principle 
be solved numerically for any value of $q$ and for an arbitrary correlation matrix ${\bf C}$. The resulting theoretical eigenvalue spectrum for the matrix $\widetilde {\bf D}$, assuming no correlation between 
the conditioning variable $\xi$ and the instantaneous correlation $x_\alpha x_\beta$, can be compared to the empirical spectrum obtained from data using Eq.~(\ref{empirical}). 
Any difference between the two spectra can be interpreted as resulting from a true correlation with the conditioning variable.

In the null-hypothesis case, it is also clear that the quantity $\widetilde \Delta$ defined by:
\be \label{deltatilde}
\widetilde \Delta = \frac{1}{\lambda_1} \left[\langle e | {\widetilde {\bf D}} | v_1 \rangle -  \langle v_1 | {\widetilde {\bf D}} | v_1 \rangle \langle e | v_1 \rangle \right].
\ee
must be zero when averaged over $\xi, x_\alpha$. One can compute its variance, which is found to be:
\be
\langle {\widetilde \Delta}^2 \rangle_{\xi,x_\alpha} = \frac{\langle e|{\bf{C}}|e \rangle - \lambda_1 \langle e| v_1 \rangle^2}{T \lambda_1} \langle \xi^2 \rangle.
\ee
For large $T$, the central limit theorem ensures that $\widetilde\Delta$ becomes Gaussian with the above variance. This result will be used below to assess whether the empirical value of $\Delta$ (defined above)
is meaningful or not.

\begin{figure}[h!btp] 
		\center
		\includegraphics[scale=0.45]{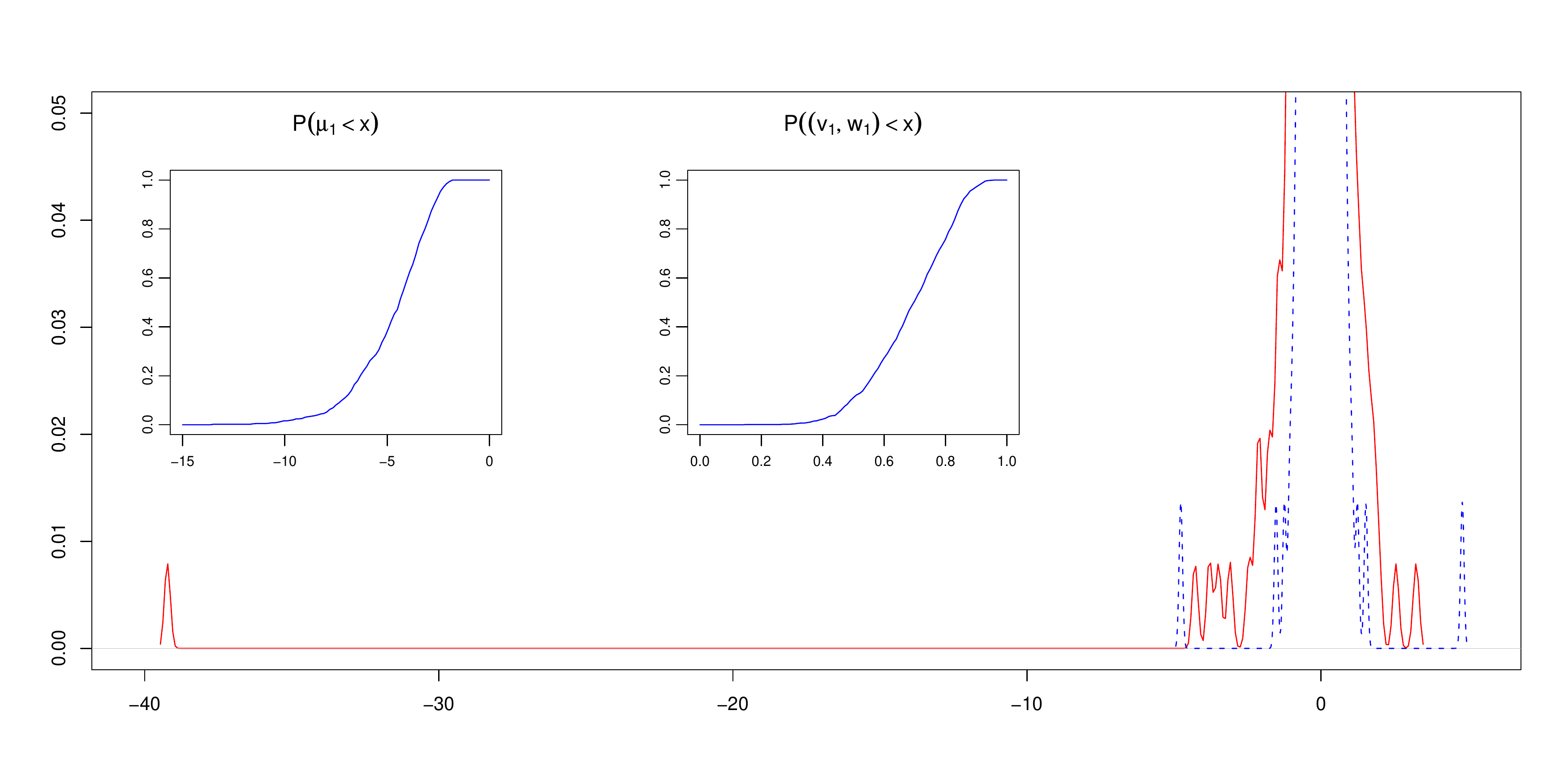}
        \caption{Main figure: empirical spectrum of ${\bf D}$ for the BE500 index (in red), compared to the null-hypothesis case (in blue). For the latter case, we have generated $1000$ 
        random samples, ranked the eigenvalues and averaged each of them separately. The leftmost blue peak therefore corresponds to the average value of the most negative eigenvalue. Insets: 
        cumulative distributions of the most negative eigenvalue $\mu_1$ and of the scalar product $S=\langle w_1|v_1\rangle$. } \label{figure3}
\end{figure}

\subsection{Numerical simulations}

In practice, however, we found it more convenient to use direct numerical simulations rather than the above exact results. In principle, these results below could be obtained using the
mathematical formalism above, but the effort required to solve numerically the equations above is larger than the one needed to make direct simulations. 
We measure the null-hypothesis spectrum of $\widetilde {\bf D}$ by choosing $\xi(t)$ to be a Gaussian random variable of zero mean and unit variance, 
completely independent of the true returns $\eta_\alpha(t)$, which we then diagonalize. The cumulative distribution of the largest negative eigenvalue in the null-hypothesis is shown in the inset. The average 
position of the most negative eigenvalue of $\widetilde {\bf D}$ in the null-hypothesis case is found to be $\widetilde{\mu}_1 \approx -4.8$. The average position of the second and third 
most negative eigenvalues in the null-hypothesis case will be denoted by $\widetilde{\mu}_2$ and $\widetilde{\mu}_3$. 

We have also measured the distribution
of the scalar product $S=\langle w_1|v_1\rangle$ between the corresponding top eigenvector $|w_1\rangle$ and the top eigenvector of ${\bf C}$, $|v_1 \rangle$. We find that even in the case 
where $\xi(t)$ is an independent random variable, the top eigenvector of ${\bf D}$ is in fact strongly correlated with $|v_1 \rangle$, with an average scalar product equal to $S=0.68$ for the correlation matrix of the
returns of the BE500 index. We find numerically that $P(S \leq 0.5) \approx 0.11$ and $P(S \leq 0.65) \approx 0.38$ for the BE500 index -- see Fig.~\ref{figure3}. Results for the SP500 are very similar.

\subsection{Comparison with empirical data}

In order to reduce the measurement noise and compare with the above numerical simulations, we have 
estimated  ${\bf D}(\tau)$ using Eq. (\ref{empirical}) with ``Gaussianized'' empirical index returns, obtained by first ranking the true index return from most negative to most positive, 
defining the rank of day $t$, $k(t)$. The Gaussianized index return $I_G(t)$ is then obtained as $\Phi^{-1}(k(t)/T)$, where $\Phi$ is the error function. 

\begin{figure}[h!btp] 
		\center
		\includegraphics[scale=0.45]{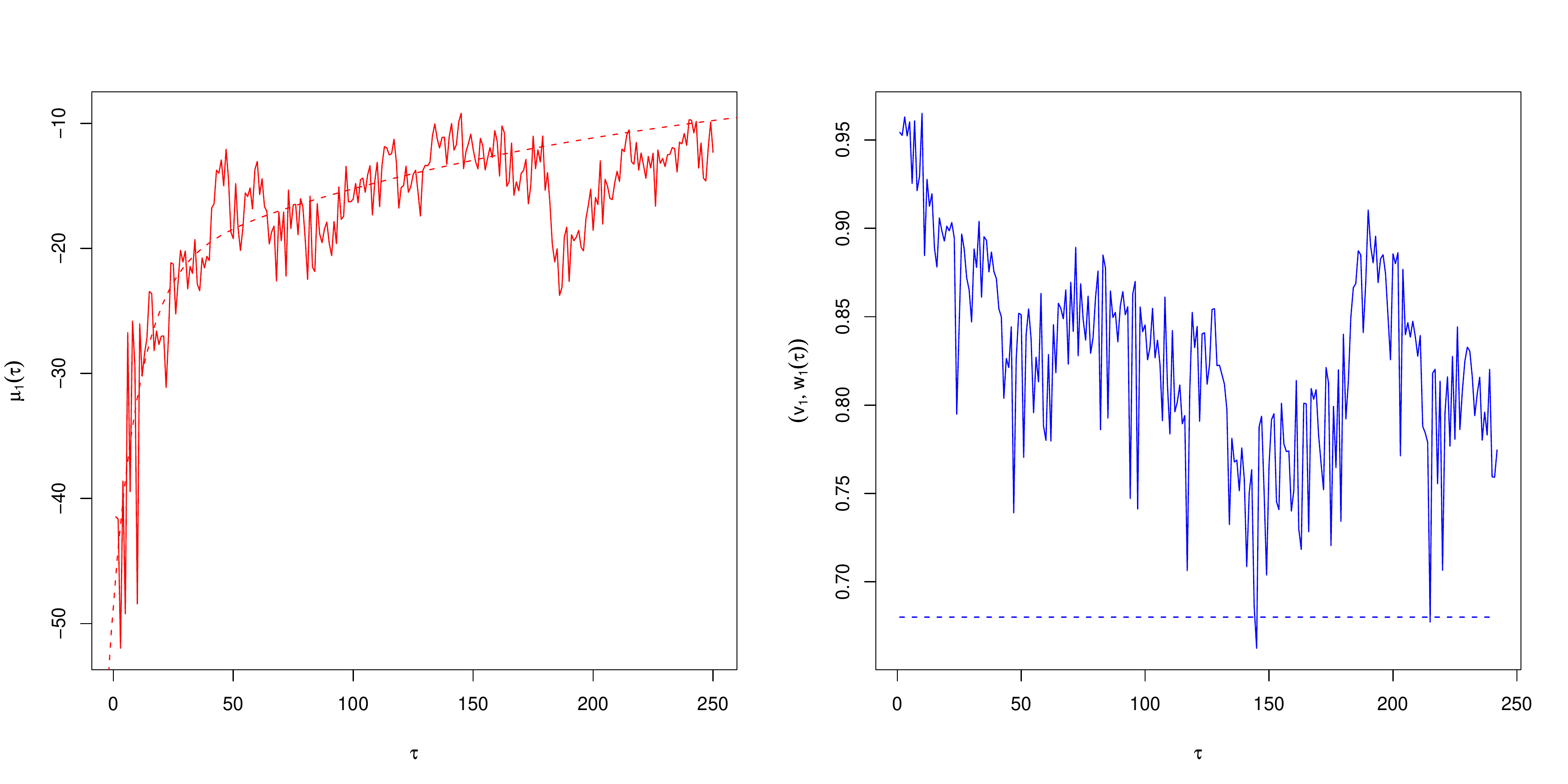}
        \caption{Left: Largest negative eigenvalue $\mu_1(\tau)$ of the lagged regression matrix ${\bf D}(\tau)$. The double exponential fit (dotted line) is given by : $\mu_1(\tau) = \mu_1^{\infty} 
        - 26.6\exp(-\tau/11) - 17.1\exp(-\tau/200)$, where we fix the value of $\mu_1^{\infty}$ using the numerical results of the previous section: $\mu_1^{\infty}=\widetilde \mu_1 \approx -4.8$,
        since we expect that for large $\tau$, all correlations are lost. 
        Right: Evolution of the scalar product $S(\tau)=\langle v_1 | w_1(\tau) \rangle$ as a function of $\tau$. The horizontal dashed line corresponds to the mean of the scalar product $S$
        in the null-hypothesis case. The data corresponds to the BE500 index, but the results for the SP500 are very similar.} \label{figure4}
\end{figure}

We show in Fig.~\ref{figure4} the evolution of $\mu_1(\tau)$, the largest (in absolute value) eigenvalue of ${\bf D}(\tau)$ as a function of $\tau$. We find that $\mu_1$ is negative, 
corresponding to the correlation leverage effect (see Eq.~(\ref{levcorr})). Comparing with the null-hypothesis case, we find that $\mu_1(\tau)$ remains significant at the $1 \%$ confidence level 
up to $\tau \approx 240$. When fitting $\mu_1(\tau)$ with an exponential function with two scales that saturates at the noise level $\widetilde \mu_1$ determined above, 
we find $\mu_1(\tau) = \widetilde \mu_1 - 26.6\exp(-\tau/11) - 17.1\exp(-\tau/200)$. This reveals two time scales; a rather short one close to the one determined directly from ${\mathcal L}_\rho(\tau)$ above 
(see Fig.~\ref{Figure1}), and a much longer time scale on the order of a year, showing that the effect of market drops on the correlation is long lasting. 
The scalar product $S(\tau)=\langle w_1(\tau) |v_1 \rangle$ between the top eigenvectors of ${\bf D}(\tau)$ and $\bf C$ globally exceeds $0.8$ in the whole 
range $\tau \in [1,240]$, whereas the null-hypothesis average value is $S=0.68$.

\begin{figure}[h!btp] 
		\center
		\includegraphics[scale=0.45]{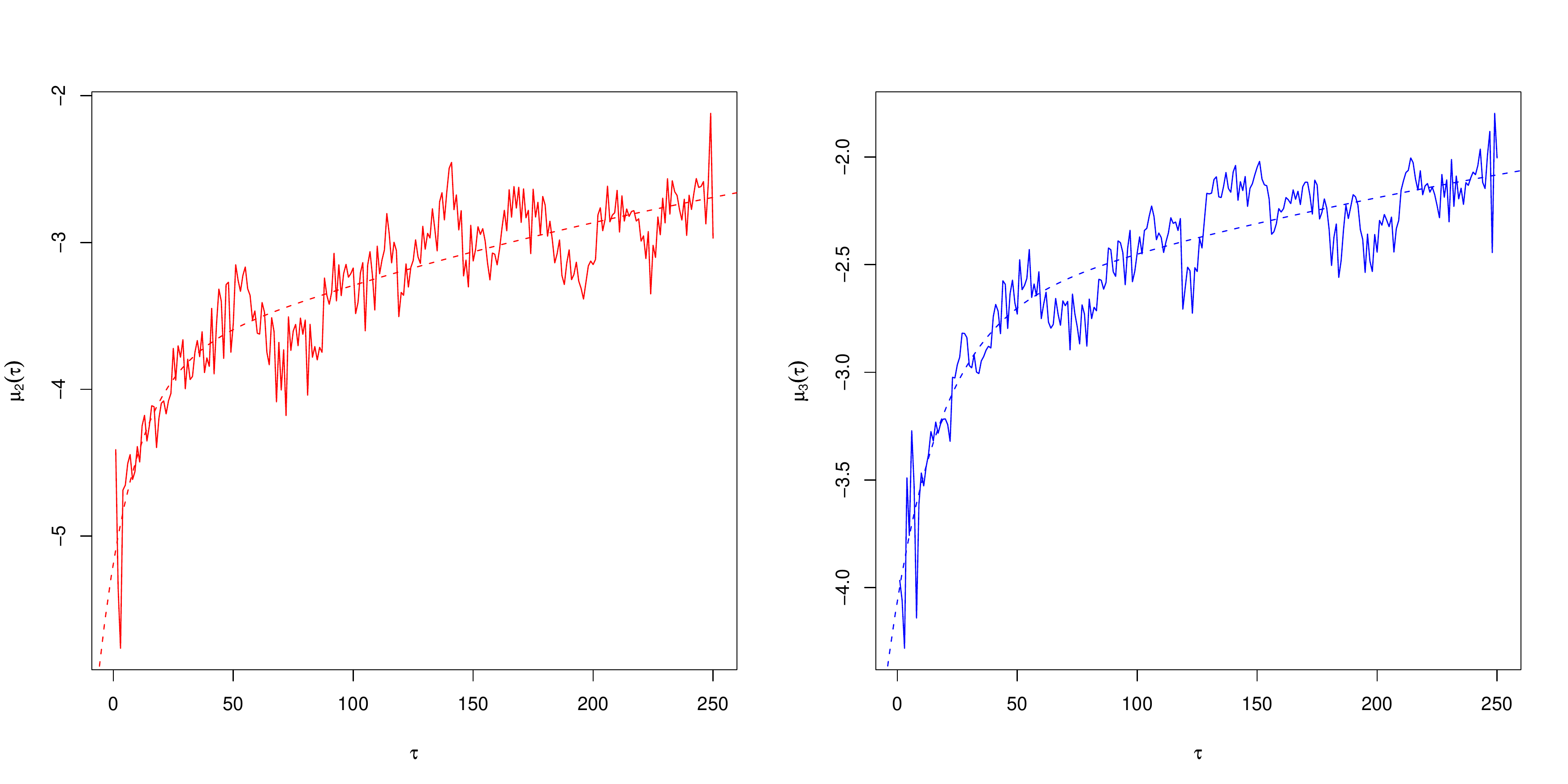}
        \caption{Left: Second eigenvalue $\mu_2(\tau)$ of the lagged regression matrix ${\bf D}(\tau)$. The exponential fit (dotted line) is given by: 
        $\mu_2(\tau)=\widetilde{\mu}_2 - 1.3 \exp(-\tau/14.4) - 2.3\exp(-\tau/364)$.
        Right: Third eigenvalue $\mu_3(\tau)$ of the lagged regression matrix ${\bf D}(\tau)$. The exponential fit (dotted line) is given by : $\mu_3(\tau)=\widetilde \mu_3 - 1.3\exp(-\tau/20) - 1.5 \exp(-\tau/420)$. 
        Direct numerical simulations of the random case lead to $\widetilde \mu_2 \approx -1.52$ and $\widetilde{\mu}_3 \approx -1.24$. 
        The data corresponds to the BE500 index, but the results for the SP500 are again very similar.}\label{figure5} 
\end{figure}

We have also studied the second ($\mu_2(\tau)$) and third ($\mu_3(\tau)$) eigenvalues of ${\bf D}(\tau)$ as a function of $\tau$, which are both negative and clearly beyond the noise level, 
and are found to decay with very similar time scales: a month and a year (see Fig.~\ref{figure5}). The corresponding eigenvectors are found to be mostly within the subspace spanned by the second and third eigenvectors of ${\bf C}$. The financial 
interpretation of these eigenvalues is of an increased sectorial correlation when the market drops on top of an increase of the market correlations. 
Therefore, all idiosyncratic effects disappear upon market drops, while global factors become dominant. 

\subsection{Separating negative \& positive returns}

\begin{figure}[h!btp] 
		\center
		\includegraphics[scale=0.45]{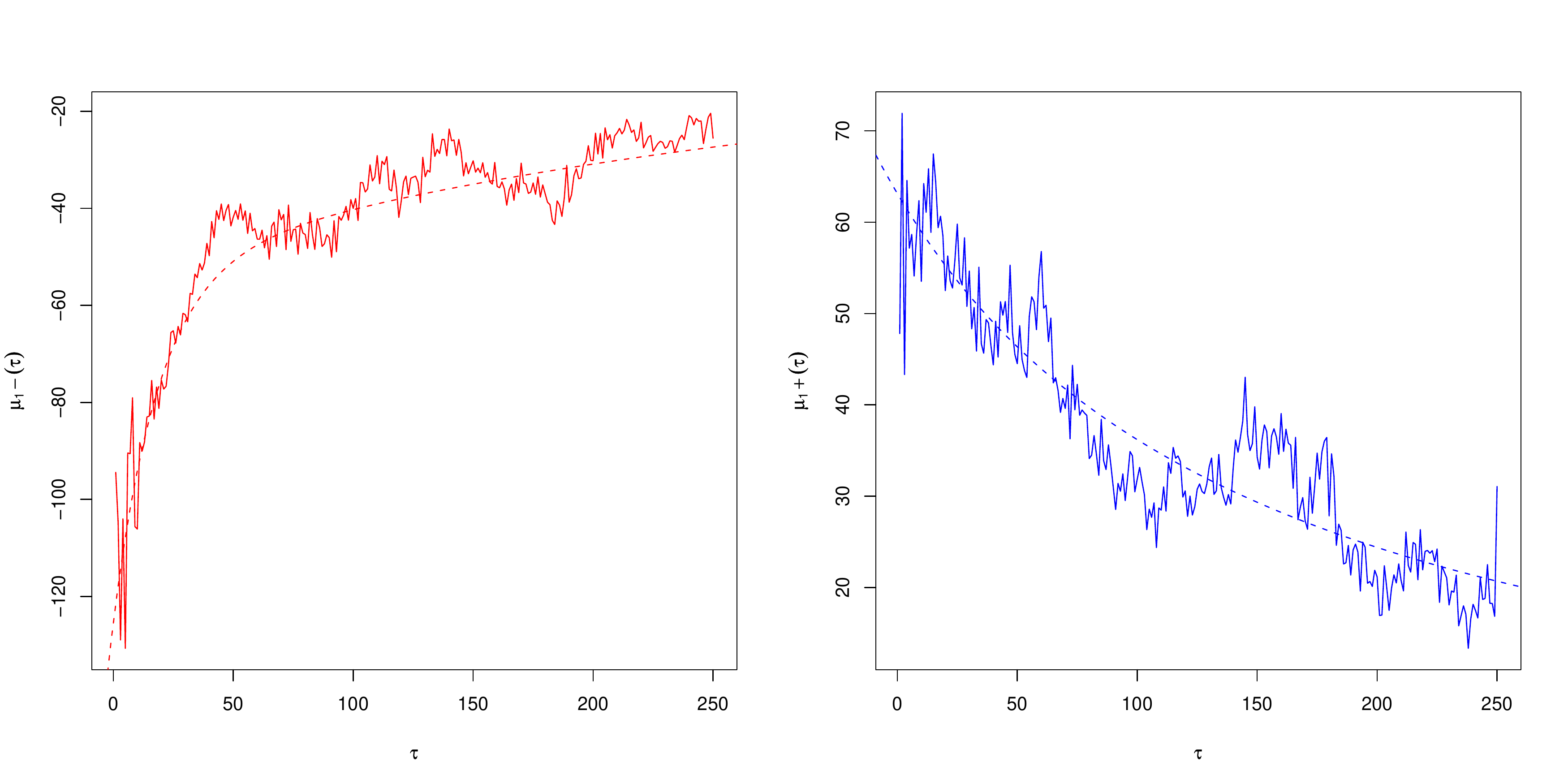}
        \caption{Left: $\mu_1^{-}(\tau)$ of the lagged regression matrix ${\bf D^{-}}(\tau)$. 
        The exponential fit (dotted line) is given by : $\mu_1^{-}(\tau)= \widetilde{\mu}_1^- - 73 \exp(-\tau/19) - 41\exp(-\tau/300)$.
        Right: $\mu_1^{+}(\tau)$ of the lagged regression matrix ${\bf D^{+}}(\tau)$. The exponential fit (dotted line) is now given by : 
        $\mu_1^{+}(\tau)=\widetilde{\mu}_1^+  + 10.6 \exp(-\tau/49) + 44 \exp(-\tau/200)$. Note again the presence of a long relaxation time on the order of a year.
        We have used direct numerical simulations to obtain $\widetilde{\mu}_1^- = \widetilde{\mu}_1^+  \approx -8.3$. The data is for the returns of the BE500. 
        Again, SP500 yields very similar results.} \label{figure6} 
\end{figure}

As Fig.~\ref{figure2} explicitely shows, the correlation depends on past index returns in a non-linear way. In fact, both negative and positive returns increase the correlations, although the effect is stronger for
negative returns, which in turn leads to a non-zero linear term in the regression of $\widehat \eta_\alpha(t)\widehat \eta_\beta(t)$ on $I(t-\tau)$. A way to capture the parabolic shape seen in Fig.~\ref{figure2}
would be to extend the above model to:
\be
\widehat \eta_\alpha(t)\widehat \eta_\beta(t) := C_{\alpha,\beta} + D_{\alpha, \beta}(\tau) I(t-\tau) + E_{\alpha, \beta}(\tau) \left[I^2(t-\tau) - \langle I^2 \rangle\right] + \varepsilon_{\alpha,\beta}(t),
\ee
defining a new matrix ${\bf E}$ that captures the symmetric effect of index returns on the correlation matrix. An alternative choice, that we adopt below, is to regress separately on negative returns and on 
positive returns:
\begin{align}
\widehat \eta_\alpha(t)\widehat \eta_\beta(t) &:= C_{\alpha,\beta} + D_{\alpha,\beta}^+(\tau) \left[I^+(t-\tau) - \langle I^+ \rangle\right] \delta_{\{I(t-\tau)>0\}} \\
&+  D_{\alpha,\beta}^-(\tau) \left[I^-(t-\tau) - \langle I^- \rangle \right] \delta_{\{I(t-\tau)<0\}} + \varepsilon_{\alpha,\beta}(t),
\end{align}
where $I^+=\max(I,0), I^-=\min(I,0)$ and $\delta$ is the Dirac function. 
With this definition, one can rewrite the correlation matrix conditioned to a certain past value of $I$ more precisely, separating the effect of positive returns and negative returns, as follows:
\be
{\bf C}(I) = {\bf C} + {\bf D^-} \left[I^- - \langle I^-\rangle \right] \delta_{\{I<0\}} + {\bf D^+} \left[I^+ - \langle I^+\rangle \right] \delta_{\{I>0\}}.
\ee
Again, in order to reduce the measurement noise, we used ``Gaussianized'' empirical index returns $I_G(t)$ instead of $I(t)$.
We apply to ${\bf D}^\pm(\tau)$ the same analysis as above. As anticipated, the top eigenvalue $\mu_1^-$ of ${\bf D}^-$ is strongly negative,
whereas the top eigenvalue $\mu_1^+$ of ${\bf D}^+$ is positive, but with $\mu_1^+ < |\mu_1^-|$ --- see Fig.~\ref{figure6}. 
The projections of $|w_1^+ \rangle$ and $|w_1^- \rangle$ onto $|v_1\rangle$ are both very close to 
unity for small $\tau$ and gradually decay to the noise level as $\tau$ increases. To check the significancy of our effect, as before, we define a null-hypothesis case, introducing the matrix: 
\begin{align}
\langle \phi^2 \rangle \widetilde{D}^-_{\alpha, \beta} = \frac{1}{T} \sum_{t=1}^T x_\alpha(t) x_\beta(t) \phi(t)
\end{align}
where the conditioning variables $\phi$ is independent of the $x_\alpha$ (which are standard gaussian variables whose correlation matrix is $\bf C$ as above) 
and distributed as $\min(\xi,0) - \langle \min(\xi,0) \rangle$ where $\xi$ is as before a standard gaussian variable. We define further the matrix $\bf \widetilde D^+$ exactly as $\bf \widetilde D^-$ except 
for the fact that the conditioning variable is now distributed as $\max(\xi,0) - \langle \max(\xi,0)\rangle$.
As above, $\widetilde{\mu}_1^-, \widetilde{\mu}_2^-, \widetilde{\mu}_3^-$ will be the average positions of the first, second and third most negative eigenvalues of $\bf \widetilde{D}^-$
and $\widetilde{\mu}_1^+, \widetilde{\mu}_2^+, \widetilde{\mu}_3^+$ will be the average positions of the first, second and third most positive eigenvalues of $\bf \widetilde{D}^+$. Those values are all computed using numerical 
simulations.  

\begin{figure}[h!btp] 
		\center
		\includegraphics[scale=0.45]{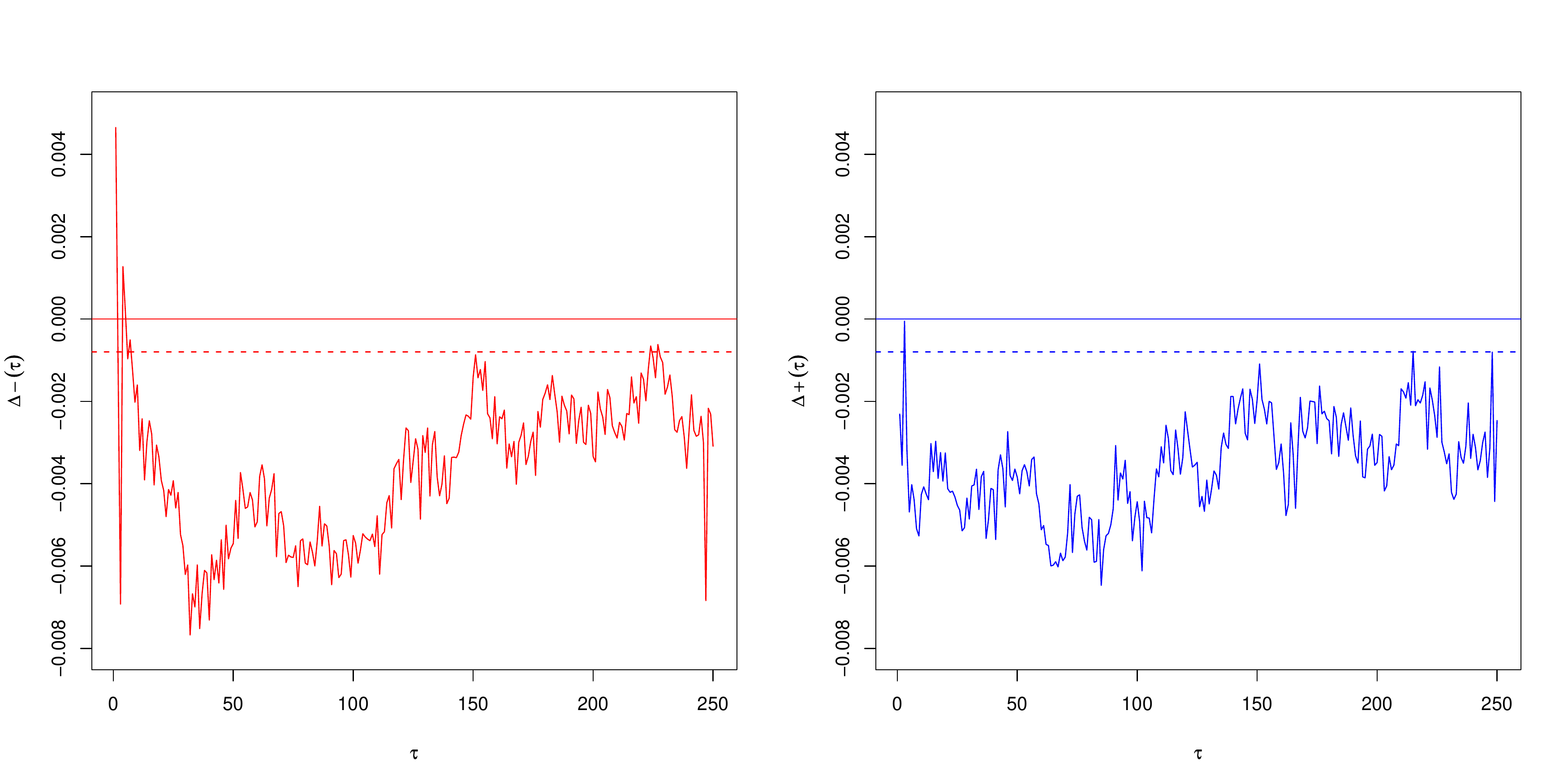}
        \caption{Plot of the rotation parameters $\Delta^-$ (left) and $\Delta^+$ (right) as a function of $\tau$. The horizontal dashed lines correspond to the 
        root-mean square error on $\Delta^\pm$ in the null-hypothesis case. The data is for BE500; the amplitude of $\Delta^-$ and $\Delta^+$ are found to be roughly a factor 2 larger for the SP500.} \label{figure7}
\end{figure}

We have also studied the rotation parameter $\Delta^\pm$ for both matrices ${\bf D}^\pm(\tau)$ defined as:
\be\label{deltapm}
\Delta^\pm = \frac{1}{\lambda_1} \left[\langle e | {\bf D}^\pm | v_1 \rangle -  \langle v_1 | {\bf D}^\pm | v_1 \rangle \langle e | v_1 \rangle \right].
\ee
The results are shown in Fig.~\ref{figure7}. In agreement with the common lore, 
$\Delta^-$ is negative, indicating that strongly negative index returns (below $\langle I^-\rangle$) lead to a more uniform instantaneous market mode. On the other hand, $\Delta^+$ is found to be negative as well, meaning that 
while strongly positive returns also tend to increase the average correlation between stocks, the instantaneous market mode rotates {\it away} from the uniform vector $| e \rangle$. The effects we are 
reporting are statistically significant since the root-mean square error on $\widetilde \Delta^\pm$ (defined as in Eq.(\ref{deltatilde})) in the null-hypothesis case is found to be $\sim 8 \cdot 10^{-4}$, a factor 3 to 4 smaller than the amplitude of 
the empirical values of $\widetilde \Delta^\pm$. 

\section{Summary \& Conclusion}

The aim of this paper was to revisit the index leverage effect, that can be decomposed into a volatility effect and a correlation effect. We investigated the latter in great detail using a matrix regression analysis, 
that we called `Principal Regression Analysis' (PRA) and for which we have provided, using Random Matrix Theory and simulations, some analytical and numerical benchmarks. 

Using this refined analysis, we confirm that downward index trends increase the average correlation between stocks (as measured by the top eigenvalue of the conditional correlation matrix), which in turn explains 
why the index leverage effect is stronger than for single stocks. Compared to the null-hypothesis benchmark, this leverage correlation effect is highly significant (see Fig.~\ref{figure4} and Fig.~\ref{figure6}). We also find 
that large downward trends implies a more uniform future market mode (see Fig.~\ref{figure7}, left).  

Upward trends, on the other hand, also increase the average correlation between stocks (see Fig.~\ref{figure6}, right) but large upward trends rotate the future market mode {\it away} from uniformity (see Fig.~\ref{figure7}, right). 
All these effects are characterized by two `memory' time scales: a `short' one on the order of a month and a longer one on the order of a year. The latter long time scale could be related to the fact that the market 
had long cycles of booms and busts within the studied time series, during which the average correlation went down and up again.

We have also studied the correlation leverage effect on intraday data, and we find (results not shown) that while the top eigenvalue of the 15 minutes correlation matrix is nearly insensitive to the sign of the previous 
15 minutes index return, a significant effect emerges when the time scale reaches one hour.  

Finally, we have found indications of a leverage effect for sectorial correlations as well, which reveals itself in the second and third modes of the PRA (see Fig.~\ref{figure5}). It would be interesting to analyze other 
conditional correlation matrices using the tools developed in this paper, such as for example leader-lagger effects \cite{Krakow,Miceli,Podobnik}, or the role of other macro variables such as oil, currencies or interest rates. 

\vskip 1cm
{\it Acknowledgements} We have benefitted from insightful comments and suggestions by Giulio Biroli, R\'emy Chicheportiche, Stefano Ciliberti, Marc Potters and Vincent Vargas. 

\bibliography{biblio_paper}

\begin{thebibliography}{10}

\bibitem{Ingve}
E.~Balogh, I.~Simonsen, B.~Nagy, and Z.~Neda.
\newblock {Persistent collective trend in stock markets}.
\newblock {\em ArXiv e-prints, 1005.0378}, 2010.

\bibitem{leverage}
G.~{Bekaert} and G.~{Wu}.
\newblock {Asymmetric volatility and risk in equity markets}.
\newblock {\em Rev. Fin. Stud.}, 13:1, 2000.

\bibitem{Bergomi1}
L.~Bergomi.
\newblock {Smile Dynamics II}.
\newblock {\em Risk}, page~67, 2005.

\bibitem{Bergomi2}
L.~Bergomi.
\newblock {Smile Dynamics III}.
\newblock {\em Risk}, page~94, 2008.

\bibitem{Biroli}
{G.} {Biroli}, {J.-P.} {Bouchaud}, and M.~{Potters}.
\newblock {The Student ensemble of correlation matrices: eigenvalue spectrum
  and Kullback-Leibler entropy}.
\newblock {\em Acta Phys. Pol. B}, 38:4009, 2007.

\bibitem{Black}
{F.} Black.
\newblock {\em {Proceedings of the 1976 American Statistical Association,
  Business and Economical Statistics Section}}.
\newblock 1976.

\bibitem{Lisanew}
L.~{Borland} and Y.~{Hassid}.
\newblock {Market panic on different time-scales}.
\newblock {\em ArXiv e-prints,1010.4917}, October 2010.

\bibitem{Miceli}
J.-P. Bouchaud, L.~Laloux, M.~A. Miceli, and M.~Potters.
\newblock {Large dimension forecasting models and random singular value
  spectra}.
\newblock {\em Eur. Phys. J.}, 201, 2007.

\bibitem{Matacz}
{J.-P.} {Bouchaud}, A.~{Matacz}, and M.~{Potters}.
\newblock {Leverage Effect in Financial Markets: The Retarded Volatility
  Model}.
\newblock {\em Physical Review Letters}, 87(22):228701--+, November 2001.

\bibitem{RMT-review}
{J.-P.} {Bouchaud} and M.~{Potters}.
\newblock {Financial Applications of Random Matrix Theory: a short review}.
\newblock {\em ArXiv e-prints}, October 2009.

\bibitem{Ciliberti}
S.~Ciliberti, J.-P. Bouchaud, and M.~Potters.
\newblock {Smile Dynamics: a Theory of the Implied Leverage Effect}.
\newblock {\em Wilmott Journal}, 1:87–94, 2009.

\bibitem{Erb}
C.B. {Erb}, C.R. {Harvey}, and T.E. {Viskanta}.
\newblock {Forecasting International Equity Correlations}.
\newblock {\em Financial Analysts Journal}, 50:32--45, 1994.

\bibitem{Glosten}
L.~Glosten, R.~Jagannathan, and D.~Runkle.
\newblock {Relationship between the expected value and the volatility of
  nominal excess return}.
\newblock {\em J. Finance}, 48:1779--1801, 1993.

\bibitem{Solnik2}
F.~Longin and B.~Solnik.
\newblock {Is the correlation in international equity returns constant:
  1960-1990}.
\newblock {\em Journal of International Money and Finance}, 14:3--26, 1995.

\bibitem{egarch}
D.~B. {Nelson}.
\newblock {Conditional Heteroskedasticity in Asset Returns: A New Approach}.
\newblock {\em Econometrica}.

\bibitem{Perello1}
J.~{Perell{\'o}} and J.~{Masoliver}.
\newblock {Random diffusion and leverage effect in financial markets}.
\newblock {\em Phys. Rev. E}, 67(3):037102--+, March 2003.

\bibitem{Perello2}
J.~{Perello}, J.~{Masoliver}, and {J.-P.} {Bouchaud}.
\newblock {Multiple time scales in volatility and leverage correlations: a
  stochastic volatility model}.
\newblock {\em Applied Mathematical Finance}, 11:27--50, 2004.

\bibitem{Podobnik}
B.~Podobnik, D.~Wang, D.~Horvatic, I.~Grosse, and H.~E. Stanley.
\newblock {Time-lag cross-correlations in collective phenomena}.
\newblock {\em Europhysics Letters}, 90, 2010.

\bibitem{Krakow}
M.~Potters, J.-P. Bouchaud, and L.~Laloux.
\newblock {Financial applications of random matrix theory : old laces and new
  pieces}.
\newblock {\em Acta Phys. Pol. B}, 36:27, 2005.

\bibitem{Ramchand}
L.~Ramchand and R.~Susmel.
\newblock {Volatility and cross-correlation across major stock markets}.
\newblock {\em Journal of Empirical Finance}, 5:397--416, 1998.

\bibitem{Solnik}
B.~Solnik, C.~Boucrelle, and Y.~Le~Fur.
\newblock {International Market Correlation and Volatility}.
\newblock {\em Financial Analysts Journal}, 52:17--34, 1996.

\bibitem{Tenenbaum}
J.~Tenenbaum, D.~Horvatic, S.C. Bajic, B.~Pehlivanovic, B.~Podobnik, and H.E.
  Stanley.
\newblock {Comparison between response dynamics in transition economies and
  developed economies}.
\newblock {\em Physical Review E}, 82:397--416, 2010.

\bibitem{Verdu}
A.~{Tulino} and S.~{Verd\`u}.
\newblock {Random Matrix Theory and Wireless Communications}.
\newblock {\em Foundations and Trends in Communication and Information Theory},
  1:1, 2004.

\end{thebibliography}
\bibliographystyle{plain}

\end{document}